\documentclass[% 
preprint, 
 amsmath,amssymb, 
 aps, 
]{revtex4} 

\usepackage{epsfig,dsfont,amssymb,amsmath,amsthm,amsfonts,amsbsy,mathrsfs} 
\usepackage{graphicx} 
\usepackage{color} 
\usepackage{bm}% bold math 
\usepackage{multirow} 
\usepackage{natbib} 

\usepackage{tikz}
\usetikzlibrary{arrows}

\newcommand{\BEQ}{\begin{equation}} 
\newcommand{\EEQ}{\end{equation}} 
\newcommand{\BEA}{\begin{eqnarray}} 
\newcommand{\EEA}{\end{eqnarray}}

\begin{document}

\title{Run-and-tumble motion in trapping environments}

\author{L. Angelani$^{1,2}$}
%\email{luca.angelani@roma1.infn.it}
\email{luca.angelani@cnr.it}

%\affiliation{$^1$ ISC-CNR, %Institute for Complex Systems, 
%Istituto dei Sistemi Complessi, P.le A. Moro 2, 00185 Rome, Italy}
\affiliation{$^1$ Istituto dei Sistemi Complessi, Consiglio Nazionale delle Ricerche, P.le A. Moro 2, 00185 Roma, Italy}
\affiliation{$^2$ Dipartimento di Fisica, Sapienza Universit\`a di Roma, P.le A. Moro 2, 00185 Roma, Italy}
%\address{$^3$  Dipartimento di Scienze Statistiche, Sapienza Universit\`a di Roma, P.le A. Moro 2, 00185 Rome, Italy}

%\address{ISC-CNR, Institute for Complex Systems, and 
%Dipartimento di Fisica, Universit\`a {\it Sapienza}, Piazzale Aldo Moro 2, 00185 Rome, Italy}

%\ead{luca.angelani@roma1.infn.it}

\begin{abstract}
Complex or hostile environments can sometimes inhibit the movement capabilities of diffusive particles or active swimmers, who may thus become stuck in fixed positions.
This occurs, for example, in the adhesion of bacteria to surfaces at the initial stage of biofilm formation.
Here we analyze the dynamics of active particles in the presence of trapping regions, where irreversible particle immobilization occurs at a fixed rate. 
By solving the kinetic equations for run-and-tumble motion in one space dimension,
we give expressions for probability distribution functions, focusing on 
stationary distributions of blocked particles, and  mean trapping times in terms of physical and geometrical parameters. Different extensions of the  trapping region are considered, from infinite to cases of semi-infinite and finite intervals.
The mean trapping time turns out to be simply the inverse of the trapping rate for infinitely extended trapping zones, while it has a nontrivial form in the semi-infinite case and is undefined for finite domains, due to the appearance of long tails in the trapping time distribution.
Finally, 
to account for the subdiffusive behavior observed in the adhesion processes of bacteria to surfaces,
we extend the model to include anomalous diffusive motion in the trapping region,
reporting the exact expression of the mean-square displacement.  
\end{abstract}

\maketitle

%%%%%%%%%%%%%%%%%%%%%%%%%%%%%%%%%%%%%%%%%%%%%%%%%%%%%%%%%%%%%%%5

\section{Introduction}
Trapping processes are quite ubiquitous in nature. Molecules can be adsorbed when they diffuse onto reactive substrates, cells can die when they move through harmful environments, living organisms can be captured by predators while foraging in hostile areas.
More recently interesting studies focused on trapping of photokinetic bacteria in structured light fields \cite{photok}.
Modeling stochastic motion in trapping environments is then of great interest \cite{Radner,Bress,Poz2021,Poz2022}. In particular, the study of active systems, composed by self-propelled particles, can give us  a very general view of the process, which applies to many interesting physical and biological phenomena \cite{Bec2016,Ecoli},
allowing diffusive motion to be obtained as a limiting case.
Understating the evolution of particles density or trapping time properties and their dependence on the physical and geometrical parameters, can give us better insights into trapping processes. In this regard, it could be very useful to determine exact expressions of these quantities in simplified models that allow analytical treatment. \\
In this paper we analyze the behavior of active particles, performing run-and-tumble motion
\cite{Sch1993,weiss,maso1992,cates2012,tai2008,mart,JPA2015,JPA2017,evans2018,mala2018}, 
in the presence of trapping regions in one space dimension. 
%%%% NEW 
%{\color{red} 
In recent years, many studies have focused on
%Many works have been devoted in the recent years to 
the analysis of trapping processes consisting of the confinement of active particles due to various causes: presence of obstacles
in crowded environments \cite{Ber2018,Riz2022,Zeit2017,Chep2013}, 
external fields or effective confining potentials induced by space-dependent motility parameters 
\cite{Bre2,Dhar2019,FPT_EPJE,Sec2019,AG2019}, 
confining boundaries \cite{Moen2022},
porous environments \cite{Bat2019,Loh2023}.
%\cite{Bre2,Dhar2019,FPT_EPJE,Sec2019,AG2019,Moen2022}.
In all these cases the term {\it trapping} indicates that the particle experiences confinement due to the presence of some kind of external cause that prevents, hinders or reduces its free motion. 
This trapping is also usually non-permanent: the particle can escape from local entrapment and continue its motion.
In this work, instead, we want to treat a different kind of {\it trapping}, and 
by this term we mean that the particle can undergo a sudden irreversible stopping of its motion when it passes through a certain region (irreversible immobilization or irreversible
trapping in a slow dynamic phase).
This is, for example, the case of bacterial adhesion to surfaces, occurring 
in the early stage of biofilms formation
\cite{Fle,Sjo,Vis,San}. 
Biofilms are complex aggregates of microorganisms that often form on surfaces and are held together by an extracellular polymeric matrix.
The complex emergent properties of this ubiquitous microbial community are of great interest from a theoretical and practical point of view.
Understanding the functional mechanisms of this ensemble of cooperating cells, involving mechanical and  physicochemical processes, 
%are still not fully understood
is not only a fascinating topic for physicists and biologists, but also an urgent task for physicians,
as biofilms are often the cause of persistent infections in living organisms \cite{Biof1}.
The first step of biofilm formation is the adhesion of cells to surfaces. 
This process, after an initial reversible phase, is essentially irreversible and the bacteria become stuck in quasi-fixed positions on surfaces.  
As a first approximation, we can therefore  describe this phenomenon as an irreversible trapping process, occurring at a given fixed rate, during the random active motion of the cell on the surface. 
We describe here such an irreversible adhesion process using a simplified one-dimensional model, which encodes the main ingredients of bacteria motion 
(run-and-tumble dynamics mimic {\it E.coli} motion \cite{Ecoli})
and trapping process (irreversible arrest).
%and allows  for analytical treatment. 
Despite its simplicity, the run-an-tumble model has been shown to capture many new and interesting phenomena of active matter, 
often allowing exact analytical expressions of many quantities of interest.
%}
%%%%%%%%%%%%%%%%%%
%{\color{red} 
In this work we use such a model to study irreversible trapping processes 
considering
%}
different extensions of trapping zones, from the simple infinite case, where the trapping region extends all over the space, to the more interesting case of semi-infinite and finite trapping zones. By solving the kinetic equations governing the evolution of probability distribution functions, we are able to obtain analytical expressions of various quantities, such as particles distributions, survival probabilities, mean-square displacements, trapping time distributions and mean-trapping times.
At the end, to account for the behaviors observed in some  experiments with bacteria, we  relax the assumption of particle immobilization  in the trapping phase and 
include the possibility of subdiffusive motion, described by fractional-type diffusion equations.\\
The paper is organized as follow. In Sec. II we define and introduce the model. In Sec. III we analyze the case of an infinitely  extended trapping region. The semi-infinite case is treated in Sec. IV and the case
of finite trapping interval is discussed in Sec. V. In Sec. VI we extend the model to describe subdiffusion
in the trapping phase, considering fractional-type equations.
Conclusions are drawn in Sec. VII.

%%%%%%%%%%%%%%%%%%%%%%%%%%%%%%%%%%%%%%%%%%%%%%%%%%%%%%%%%%%%%%%5
\section{Run-and-tumble model in trapping regions}

We consider a  run-and-tumble particle moving at constant speed $v$ and 
reorienting its direction of motion with rate $\alpha$.
We are interested in describing the particle motion when immersed in
absorbing environments which cause the irreversible trapping of the particle with a given rate $\gamma(x)$, which, in general, is a space-dependent quantity.
We denote with  $P_{_R}(x,t)$ and  $P_{_L}(x,t)$ the probability density functions (PDF)
of right-oriented and left-oriented active (moving) particles and with $P_{_B}(x,t)$ 
the PDF of blocked (trapped, immobilized) particles.
The general equations describing the time evolution of the PDFs are
\begin{eqnarray}
\label{eq_genR}
\frac{\partial P_{_R}}{\partial t}(x,t) &= -& v \frac{\partial P_{_R} }{\partial x}(x,t)
- \frac{\alpha}{2} P_{_R}(x,t) + \frac{\alpha}{2} P_{_L}(x,t) - \gamma(x) P_{_R}(x,t) , \\
\label{eq_genL}
\frac{\partial P_{_L}}{\partial t}(x,t) &= & v \frac{\partial P_{_L} }{\partial x}(x,t)
- \frac{\alpha}{2} P_{_L}(x,t) + \frac{\alpha}{2} P_{_R}(x,t) - \gamma(x) P_{_L}(x,t) ,\\
\label{eq_genB}
\frac{\partial P_{_B}}{\partial t}(x,t) &= & \gamma(x) [P_{_R}(x,t) + P_{_L}(x,t)] .
\end{eqnarray}
For $\gamma=0$ the first two equations reduce to the standard equations describing run-and-tumble
particles 
\cite{Sch1993,weiss,maso1992,cates2012,tai2008,mart,JPA2015,JPA2017,evans2018,mala2018}.
In the following we analyze different cases, from the homogeneous one in which a particle
moves in a infinitely extended trapping region 
($\gamma$ is constant throughout the whole space)
to more complex heterogeneous situations in which
the trapping zones have finite or semi-infinite extension ($\gamma$ is space dependent step function).
In all the investigated cases we will consider a particle that symmetrically starts its motion at the origin, $P_{_R}(x,0)=P_{_L}(x,0)=\delta(x)/2$, and it is immersed in a symmetric environment, i.e., $\gamma(-x)=\gamma(x)$.

%%%%%%%%%%%%%%%%%%%%%%%%%%%%%%%%%%%%%%%%%%%%5

\section{Infinitely extended  trapping region}
\label{sec3}

\begin{figure}[t!]
\includegraphics[width=.8\linewidth] {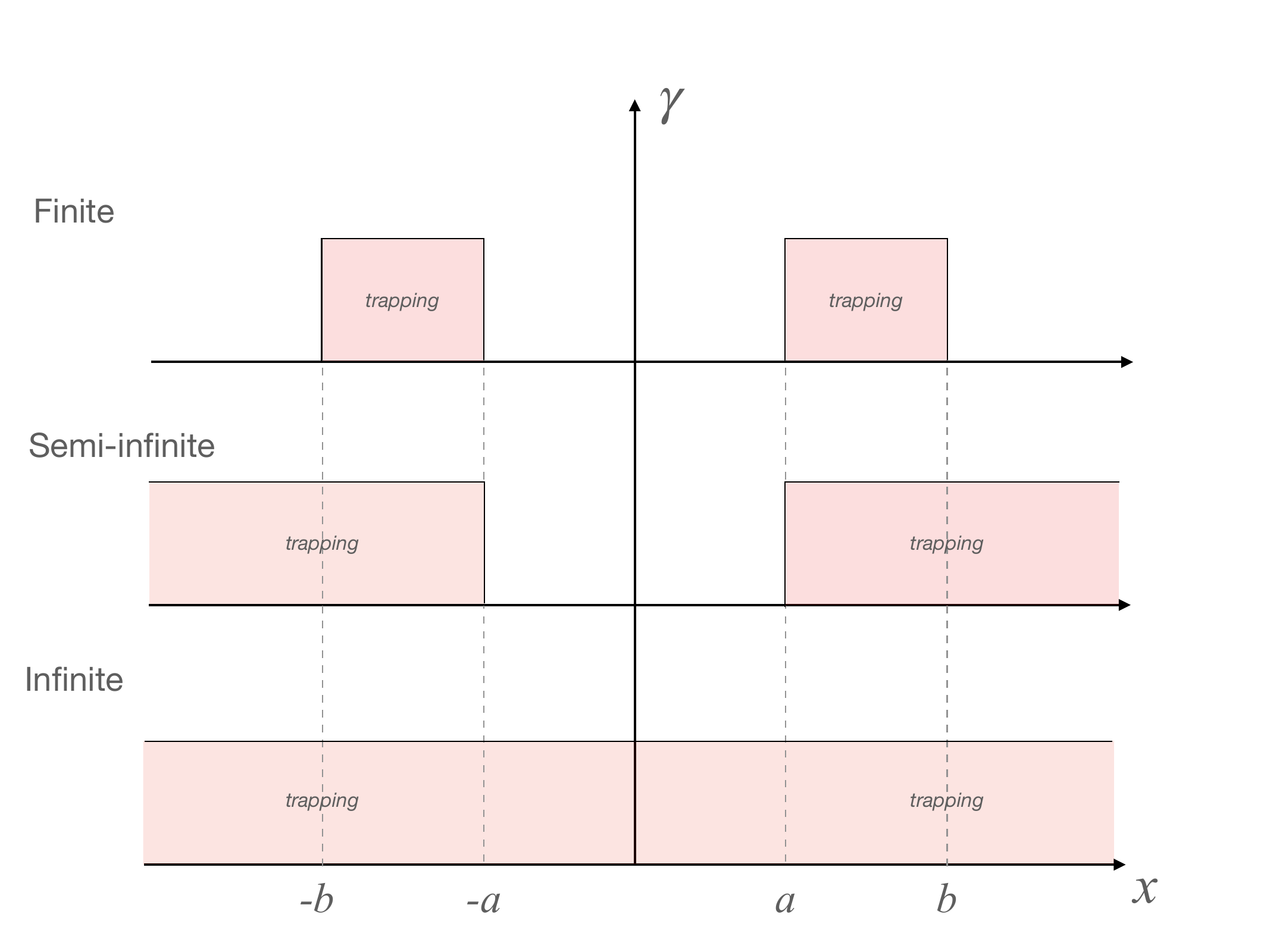}
\caption{\label{fig1}
Sketch of the trapping zones in the three cases analyzed in this study, corresponding to 
infinite, semi-infinite and finite extensions of the intervals 
where an irreversible immobilization of particles occurs at rate $\gamma$.
}
\end{figure}

We first consider the case of a run-and-tumble particle 
moving in a infinitely extend trapping region (see Fig.\ref{fig1}).
The model is described by the following equations with non-vanishing and constant $\gamma(x)=\gamma$ (for the sake of simplicity we do not indicate the dependence on space and time variables)
\begin{eqnarray}
\label{eq_r}
\frac{\partial P_{_R}}{\partial t} &= -& v \frac{\partial P_{_R} }{\partial x} 
- \frac{\alpha}{2} P_{_R} + \frac{\alpha}{2} P_{_L} - \gamma P_{_R} ,\\
\label{eq_l}
\frac{\partial P_{_L}}{\partial t} &= & v \frac{\partial P_{_L} }{\partial x} 
- \frac{\alpha}{2} P_{_L} + \frac{\alpha}{2} P_{_R} - \gamma P_{_L} ,\\
\label{eq_b}
\frac{\partial P_{_B}}{\partial t} &= & \gamma ( P_{_R} + P_{_L}) .
\end{eqnarray}
By defining the PDF of active particles $P=P_{_R} + P_{_L}$ and
the current $J=v(P_{_R} - P_{_L})$ we have
\begin{eqnarray}
\label{eq_PA}
\frac{\partial P}{\partial t} &= & -\frac{\partial J }{\partial x} - \gamma P ,\\
\label{eq_J}
\frac{\partial J}{\partial t} &= & - v^2 \frac{\partial P }{\partial x} 
- (\alpha+ \gamma) J  ,\\
\label{eq_PB}
\frac{\partial P_{_B}}{\partial t} &= & \gamma P ,
\end{eqnarray}
with the normalization $\int dx (P+P_{_B}) = 1$.
By using the Laplace transform
\begin{equation}
{\tilde P} (s) \equiv
    \mathcal{L}[P(t)] (s)=
\int_0^\infty dt  \ e^{-st} \ P(t) ,
\end{equation}
and considering initial conditions
$P(x,0) = \delta(x)$, 
$J(x,0) = 0$, $P_{_B}(x,0) = 0$,
we have
\begin{eqnarray}
\label{eq_PA_L}
\frac{\partial {\tilde{J}}}{\partial x} &=& \delta(x) - (s+\gamma) {\tilde P} ,\\
\label{eq_J_L}
v^2 \frac{\partial {\tilde P}}{\partial x} &=& - (s+\alpha+ \gamma) {\tilde J} ,\\
\label{eq_PB_L}
{\tilde P}_{_B} &=&  \frac{\gamma}{s} {\tilde P} .
\end{eqnarray}
By combining the first two equations we obtain the second order differential equation for ${\tilde P}$
\begin{equation}
v^2 \frac{\partial^2 {\tilde P}}{\partial x^2} - 
(s+\gamma) (s+\alpha+ \gamma) {\tilde P} = 
- (s+\alpha+ \gamma)\ \delta(x) ,
\label{teleg}
\end{equation}
whose solution is 
\begin{equation} 
\label{PA_L}
{\tilde P}(x,s) = \frac{1}{2v} \sqrt{\frac{s+\alpha+\gamma}{s+\gamma}} \ \exp{\left(-c|x|\right)} ,
\end{equation} 
where $c(s)$ is
\begin{equation}
v^2 c^2 = (s+\gamma)(s+\alpha+\gamma) .
\end{equation}

\noindent
We note that the active particle PDF (\ref{PA_L}) is the Laplace-shifted solution of the classical
one-dimensional PDF of a run-and-tumble particle in free space $P_0$,
${\tilde P}(x,s) = {\tilde P}_0(x,s+\gamma)$,
as it is also evident by noting that Eq.s (\ref{eq_PA_L},\ref{eq_J_L}) and (\ref{teleg}) are identical to those of standard run-and-tumble particles in free (non-trapping) space with the substitution $s \to s+\gamma$ (see Eq.s (15,16) of Ref.\cite{JPA2015}).
Therefore, in the time domain, we have that 
$P(x,t) = \exp{(-\gamma t)}P_0(x,t)$.
The free solution $P_0(x,t)$ is well known in the literature 
(see, for example, \cite{weiss,mart})
and then we can write the explicit expression of $P(x,t)$ as
\begin{eqnarray}
P(x,t) & = &  \frac{e^{-(\gamma+\alpha/2) t}}{2} \Biggl\{
\delta(x-vt) + \delta(x+vt) \Biggr. \nonumber \\
 & + & \left.
\left[
\frac{\alpha}{2v}\ I_0\left(\frac{\alpha \Delta(x,t)}{2v}\right) \right. \right.
 + \Biggl. \left. 
\frac{\alpha t}{2 \Delta(x,t)}\  I_1\left(\frac{\alpha \Delta(x,t)}{2v}\right) \right] \theta(vt-|x|)
\Biggr\} ,
\label{PAxt}
\end{eqnarray}
where $I_0$, $I_1$  are modified Bessel functions and
%\begin{equation}
$\Delta=\sqrt{v^2t^2- x^2}$.
%\end{equation}
\begin{figure}[t!]
\includegraphics[width=.44\linewidth] {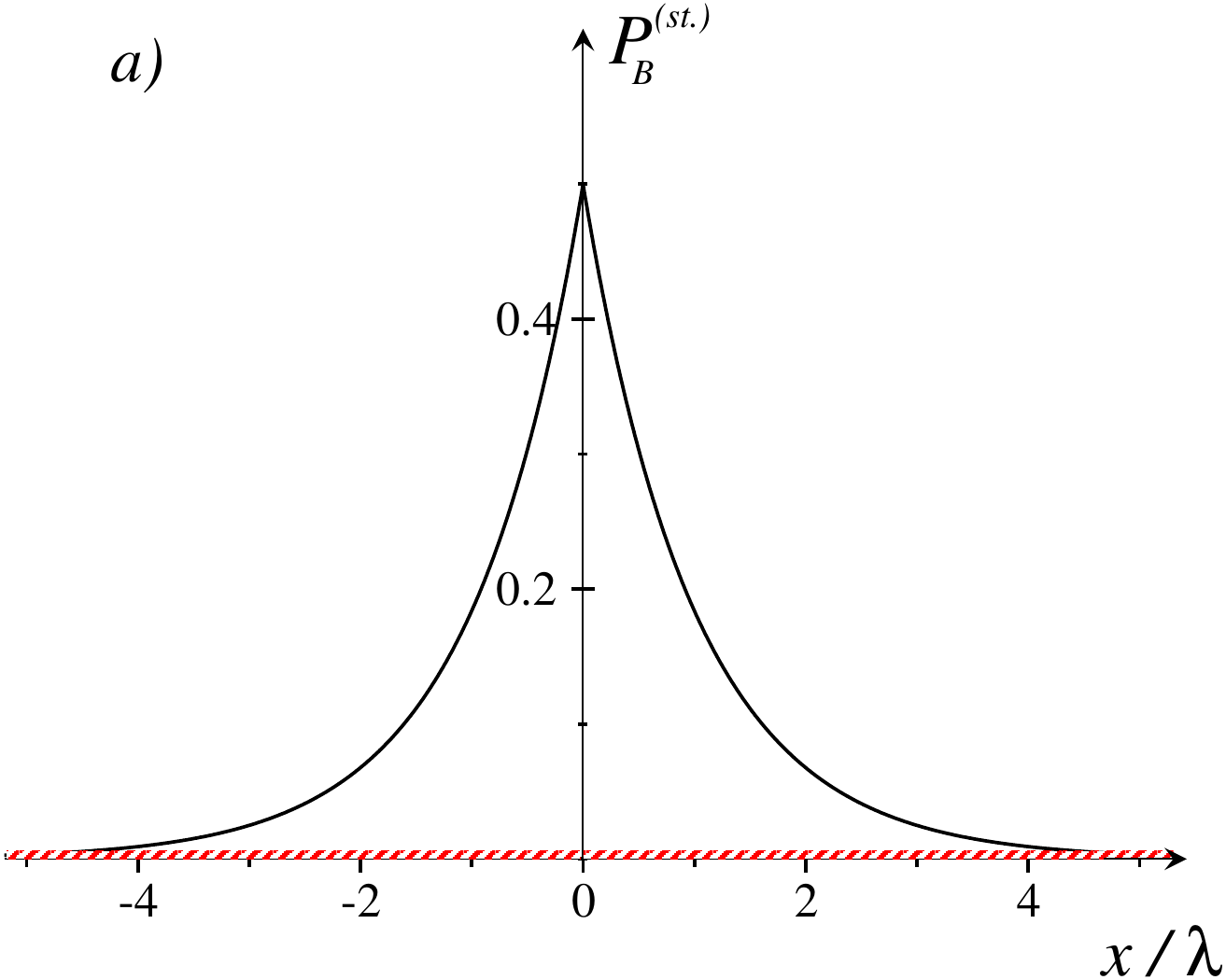}
\includegraphics[width=.44\linewidth] {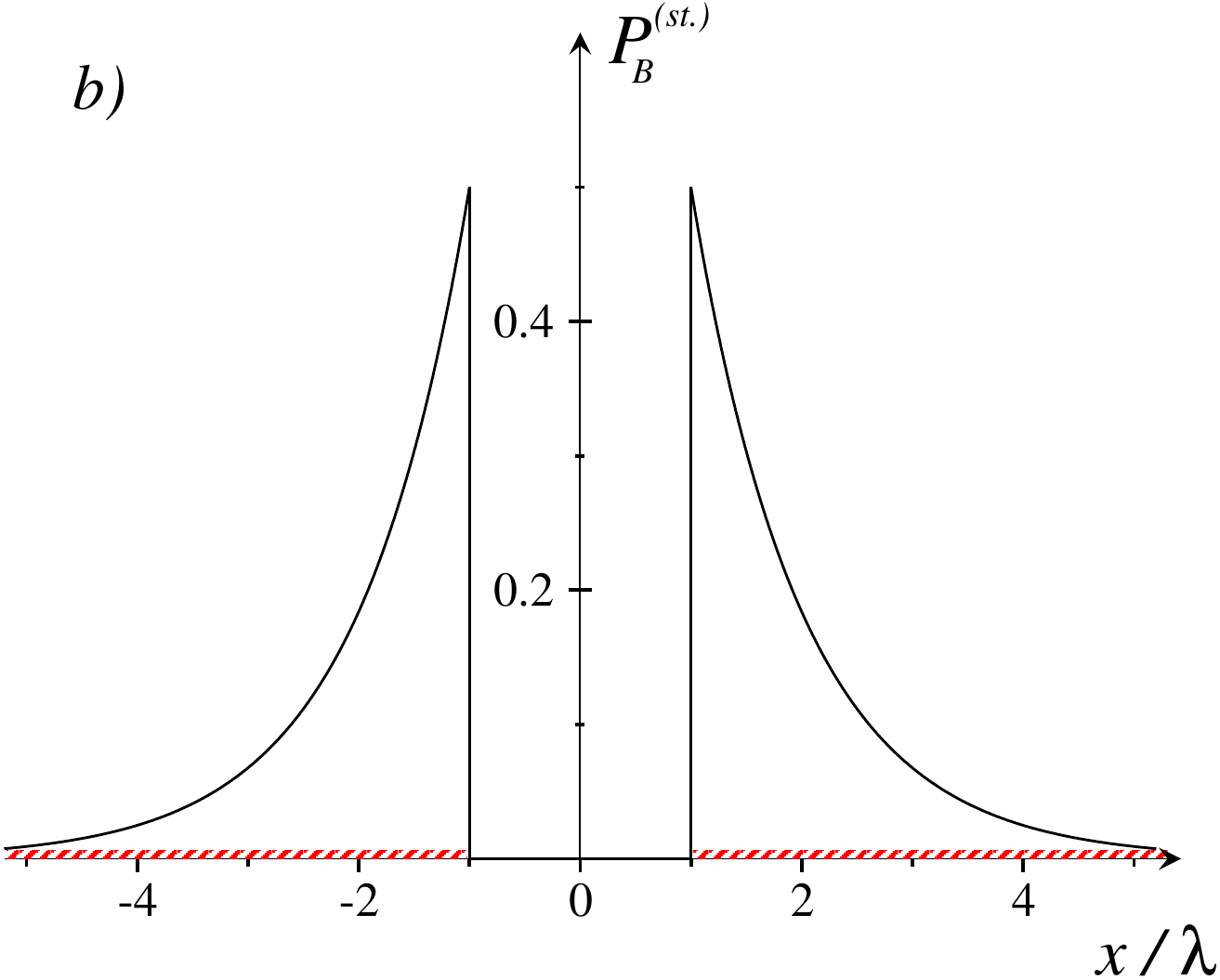}
\includegraphics[width=.44\linewidth] {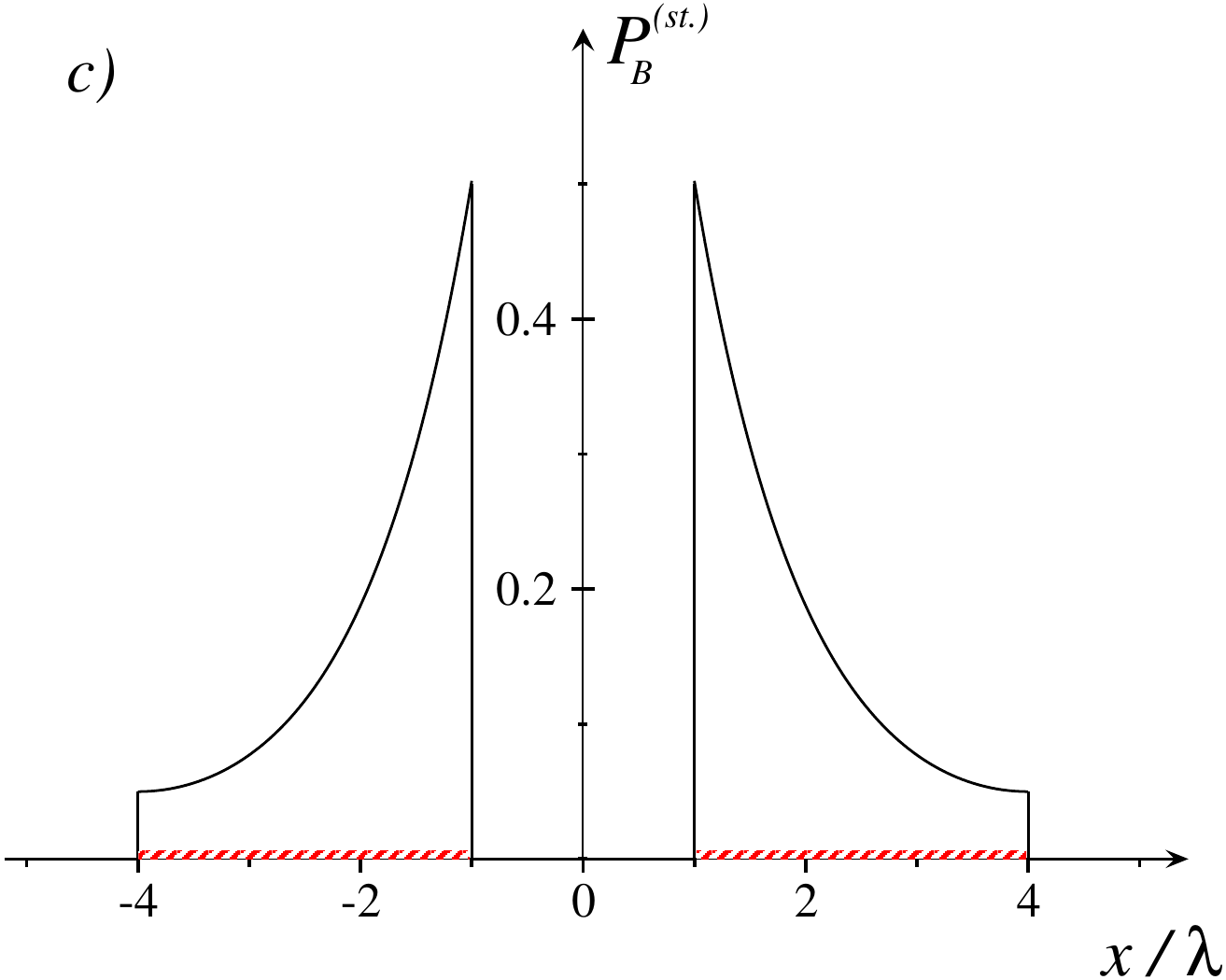}
\caption{\label{fig2}
Stationary probability distributions $P_{_B}^{(st.)}(x)$
of blocked particles
in trapping regions (where $\gamma>0$, highlighted  red zones along the $x$ axis). 
{\it a}) The trapping zone extends all over the space, Eq.(\ref{PBst1}).
{\it b}) The case of semi-infinite trapping regions, $|x|> a$, with $a/\lambda=1$, Eq.(\ref{PBst2}).
{\it c}) The case of finite trapping intervals, $a<|x|<b$, with $a/\lambda=1$ and $b/\lambda=4$, Eq.(\ref{PBst3}). We set $\alpha=1$, $v=1$ and $\gamma=1$.
}
\end{figure}
The PDF of blocked particles is obtained as time-integral of $P$, being their Laplace transforms related
through Eq.(\ref{eq_PB_L}):
\begin{equation}
    P_{_B}(x,t) = \gamma \ \int_0^t dt' \ P(x,t') .
\end{equation}
After some algebra we finally obtain
\begin{eqnarray}
P_{_B}(x,t) & =&  \frac{\gamma}{2v} \left[ 
e^{-(\gamma+\alpha/2) t}  I_0\left(\frac{\alpha \Delta(x,t)}{2v}\right) \right. \nonumber \\
& + & \left. (\gamma+\alpha)
\int_{|x|/v}^t dt' \ e^{-(\gamma+\alpha/2) t'}
  I_0\left(\frac{\alpha \Delta(x,t')}{2v}\right) \right] \theta(vt-|x|) .
\end{eqnarray}
The stationary distribution of blocked particles is given by
$P_{_B}^{(st.)}(x) = \lim_{t \to \infty} P_{_B}(x,t) =
\lim_{s \to 0} s {\tilde P}_{_B}(x,s)$, leading to
\begin{equation}
\label{PBst1}
P_{_B}^{(st.)} (x) = \frac{1}{2\lambda} \ \exp{\left(-\frac{|x|}{\lambda}\right)} ,
\end{equation}
where we have introduced the characteristic length $\lambda$ 
\begin{equation}
\lambda = \frac{v}{\sqrt{\gamma (\alpha+\gamma)}} .
\label{lamb}
\end{equation}
Fig.\ref{fig2} shows the stationary distribution (\ref{PBst1}), along with those for semi-infinite and finite trapping zones (see following sections).
We note that, in the diffusive limit, $\alpha,v \to \infty$ with finite 
diffusion constant $D=v^2/\alpha$,
the characteristic length reads $\lambda_{Diff.} = \sqrt{D/\gamma}$ and the stationary distribution
(\ref{PBst1}) reduces to that obtained in \cite{Kos2023}:
\begin{equation}
P_{_{B,Diff.}}^{(st.)} (x) = \frac{1}{2}\sqrt{\frac{\gamma}{D}} \ 
\exp{\left(- \sqrt{\frac{\gamma}{D}} |x| \right)} .
\end{equation}
We now study the probability distribution $\varphi(t)$ of the trapping time, which is related to the survival probability  $\mathbb{P}(t)$ , i.e., the probability that the active particle 
has not been trapped until time $t$
\begin{equation}
    \varphi(t) = - \frac{\partial \mathbb{P}}{\partial t} (t) .
\end{equation}
The survival probability is obtained as an integration over  space of the active particles PDF
\begin{equation}
    \mathbb{P}(t)= \int_{-\infty}^\infty dx\ P(x,t) .
    \label{SurvP}
\end{equation}
By using (\ref{PA_L}), the Laplace transform is given by
\begin{equation}
    {\tilde {\mathbb{P}}}(s) = \frac{1}{s+\gamma} ,
\end{equation}
corresponding, in the time domain, to 
\begin{equation}
\mathbb{P}(t)= \exp{(-\gamma t)} .
\end{equation}
The trapping times are then exponentially distributed 
\begin{equation}
\label{phi_inf}
    \varphi(t) = \gamma \exp{(-\gamma t)} ,
\end{equation}
and the mean trapping time 
\begin{equation}
\tau = \int_0^\infty dt \ t \ \varphi(t) ,
\label{MTT}
\end{equation}
is simply the inverse of the trapping rate
\begin{equation}
\tau = \frac{1}{\gamma} .
\end{equation}
Another interesting quantity to calculate is the mean-square displacement (MSD) of particles, i.e. the second moment of the total particle distribution function $P+P_{_B}$
\begin{equation}
    r^2(t) = \int_{-\infty}^\infty dx \  x^2 \ [P(x,t) + P_{_B}(x,t)] .
\end{equation}
Working in the Laplace domain, using  (\ref{PA_L}) and (\ref{eq_PB_L}), we have
\begin{equation}
\label{r2Lap}
    {\tilde{r^2}}(s) = \frac{2 v^2}{s (s+\gamma)(s+\gamma+\alpha)} .
\end{equation}
Inverting the Laplace transform we finally obtain the expression of the MSD 
\begin{equation}
r^2(t) = \frac{2v^2}{\alpha \gamma (\alpha+\gamma)} 
\left[ 
\alpha (1-e^{-\gamma t}) - 
\gamma e^{-\gamma t } (1-e^{-\alpha t})
\right] .
\label{r2}
\end{equation}
\begin{figure}[t!]
\includegraphics[width=.7\linewidth] {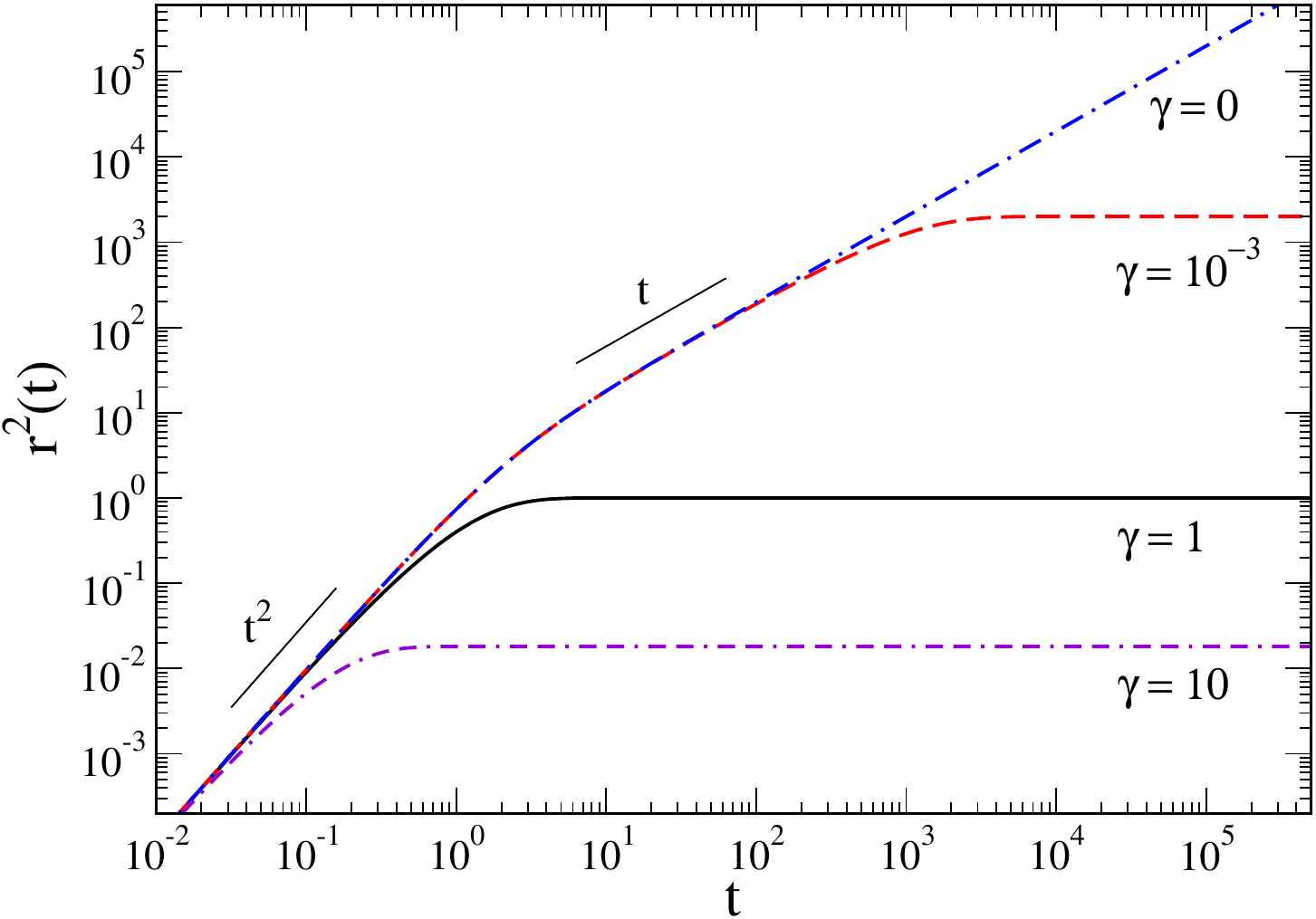}
\caption{\label{fig3}
Mean-square displacement $r^2(t)$ in the case of infinitely extended trapping region for different 
values of the trapping rate: $\gamma=0$ (absence of trapping), $\gamma=10^{-3}$,
$\gamma=1$ and $\gamma=10$. One observes ballistic behavior $r^2 \sim t^2$ at short times ($t \ll \min{(\alpha^{-1},\gamma^{-1})}$), possibly
diffusive one $r^2 \sim t$ at intermediate times ($\alpha^{-1}<t<\gamma^{-1}$) and
saturation $r^2 \to r^2_{\infty} $ at long times ($t \gg \max{(\alpha^{-1},\gamma^{-1})}$).
We set $\alpha=1$ and $v=1$.
}
\end{figure}
\noindent We note that, for $\gamma=0$, the above expression reduces to the usual one for run-and-tumble free particles \cite{weiss}
\begin{equation}
r^2(t) = \frac{2v^2}{\alpha^2} 
\left[
\alpha t -1 + e^{-\alpha t} \right] ,
\hspace{1.3cm} \gamma \to 0.
\end{equation}
The asymptotic limit of (\ref{r2}) is finite
\begin{equation}
\label{r2plat}
    r_\infty^2 = \frac{2v^2}{\gamma(\alpha+\gamma)} ,
    \hspace{1.3cm} t \to \infty,
 \end{equation}
which is, indeed, the second moment of the blocked particles distribution in the stationary regime (\ref{PBst1}).
We finally observe that, in the diffusive limit, the MSD reads
\begin{equation}
r^2_{Diff.}(t) = \frac{2D}{\gamma} 
\left( 1-e^{-\gamma t} \right) ,
\hspace{1.3cm} v,\alpha \to \infty \text{ with } D=v^2/\alpha ,
\label{r2diff}
\end{equation}
which, for $\gamma \to 0$, reduces to the standard form in the free space 
\begin{equation}
r^2_{Diff.}(t) = 2Dt , 
\hspace{1.3cm} \gamma \to 0.
\end{equation}
In Fig.\ref{fig3} the MSD (\ref{r2}) is shown for four different values of the trapping parameter
$\gamma$.

%%%%%%%%%%%%%%%%%%
\section{Semi-infinite trapping region}
We now consider the case in which the trapping region is $|x|>a$
(see Fig.\ref{fig1}).
We have to solve two sets of Eq.s(\ref{eq_genR}-\ref{eq_genB}). 
In the free region (I) $|x|<a$ we have $\gamma=0$:
\begin{eqnarray}
\label{PR_I}
\frac{\partial P^{^{({\rm I})}}_{_R}}{\partial t} &= -& v \frac{\partial 
P^{^{({\rm I})}}_{_R} }{\partial x} 
- \frac{\alpha}{2} P^{^{({\rm I})}}_{_R} + \frac{\alpha}{2} P^{^{({\rm I})}}_{_L} ,\\
\label{PL_I}
\frac{\partial P^{^{({\rm I})}}_{_L}}{\partial t} &= & v \frac{\partial P^{^{({\rm I})}}_{_L} }{\partial x} 
- \frac{\alpha}{2} P^{^{({\rm I})}}_{_L} + \frac{\alpha}{2} P^{^{({\rm I})}}_{_R} .
\end{eqnarray}
In the trapping region (II) $|x|>a$ we have $\gamma>0$:
\begin{eqnarray}
\label{PR_II}
\frac{\partial P^{^{({\rm II})}}_{_R}}{\partial t} &= -& v \frac{\partial 
P^{^{({\rm II})}}_{_R} }{\partial x} 
- \frac{\alpha}{2} P^{^{({\rm II})}}_{_R} + \frac{\alpha}{2} P^{^{({\rm II})}}_{_L} - \gamma P^{^{({\rm II})}}_{_R} ,\\
\label{PL_II}
\frac{\partial P^{^{({\rm II})}}_{_L}}{\partial t} &= & v \frac{\partial P^{^{({\rm II})}}_{_L} }{\partial x} 
- \frac{\alpha}{2} P^{^{({\rm II})}}_{_L} + \frac{\alpha}{2} P^{^{({\rm II})}}_{_R} - \gamma P^{^{({\rm II})}}_{_L} ,\\
\frac{\partial P_{_B}}{\partial t} &= & \gamma ( P^{^{({\rm II})}}_{_R} + P^{^{({\rm II})}}_{_L}) .
\label{PB_II}
\end{eqnarray}
We note that blocked particles are present only in the trapping region (II).
The corresponding differential equations for the probability density $P=P_{_R}+P_{_L}$ in the two zones are then given by (\ref{teleg}) with $\gamma=0$ in the free zone (I) and with $\gamma>0$ in the trapping zone (II).
By imposing continuity condition for $P$ and discontinuity for $\partial_x P$ in $|x|=a$ 
(continuity of the current $J$) we finally obtain, in the Laplace domain
\begin{eqnarray}
\label{sPA_I}
{\tilde P}^{^{({\rm I})}}(x,s) &=& A^{^{({\rm I})}}_+\ \exp{\left(c_{_0}|x|\right)}
+ A^{^{({\rm I})}}_-\ \exp{\left(-c_{_0}|x|\right)},
\quad \mbox{for $|x|<a$},\\
\label{sPA_II}
{\tilde P}^{^{({\rm II})}}(x,s) &=& 
A^{^{({\rm II})}}_-\ \exp{\left(-c|x|\right)} ,
\quad \mbox{for $|x|>a$},\\
{\tilde P}_{_B} (x,s) &=&  \frac{\gamma}{s} 
{\tilde P}^{^{({\rm II})}}(x,s) ,
\quad \mbox{for $|x|>a$} ,
\label{sPB_II}
\end{eqnarray}
where
\begin{eqnarray}
\label{cs}
v^2 c^2 &=& (s+\gamma)(s+\alpha+\gamma) ,\\
v^2 c_{_0}^2 &=& s (s+\alpha) ,
\label{c0s}
\end{eqnarray}
and 
\begin{eqnarray}
A^{^{({\rm I})}}_{\pm} &=& \mp  \frac{c_{_0}}{4s} \ 
\frac{c \mp c_{_0}q}{c \cosh{(c_{_0}a)} + c_{_0} q \sinh{(c_{_0} a)}} \ \exp{(\mp c_{_0} a)} ,\\
A^{^{({\rm II})}}_{-} &=& \frac{c_{_0}^2 q}{2s} \ 
\frac{1}{c \cosh{(c_{_0}a)} + c_{_0} q \sinh{(c_{_0} a)}} ,
\ \exp{(c a)} 
\end{eqnarray}
with 
\begin{equation}
    q = \frac{s+\alpha+\gamma}{s+\alpha} .
\label{qs}
\end{equation}
The stationary distribution of blocked particles is
obtained as $P_{_B}^{(st.)}(x) = \lim_{s \to 0} s {\tilde P}_{_B}(x,s) = 
\lim_{s \to 0} \gamma {\tilde P}^{^{({\rm II})}}((x,s)$, giving rise to (see Fig.\ref{fig2})
\begin{equation}
\label{PBst2}
P_{_B}^{(st.)} (x) = \frac{1}{2\lambda} \ \exp{\left(-\frac{|x|-a}{\lambda}\right)},
\quad \mbox{for $|x|>a$},
\end{equation}
with $\lambda$ given by ({\ref{lamb}}).
For $a=0$ the above expression reduces to (\ref{PBst1}), valid in the case of infinite trapping region.\\
\begin{figure}[t!]
\includegraphics[width=.7\linewidth] {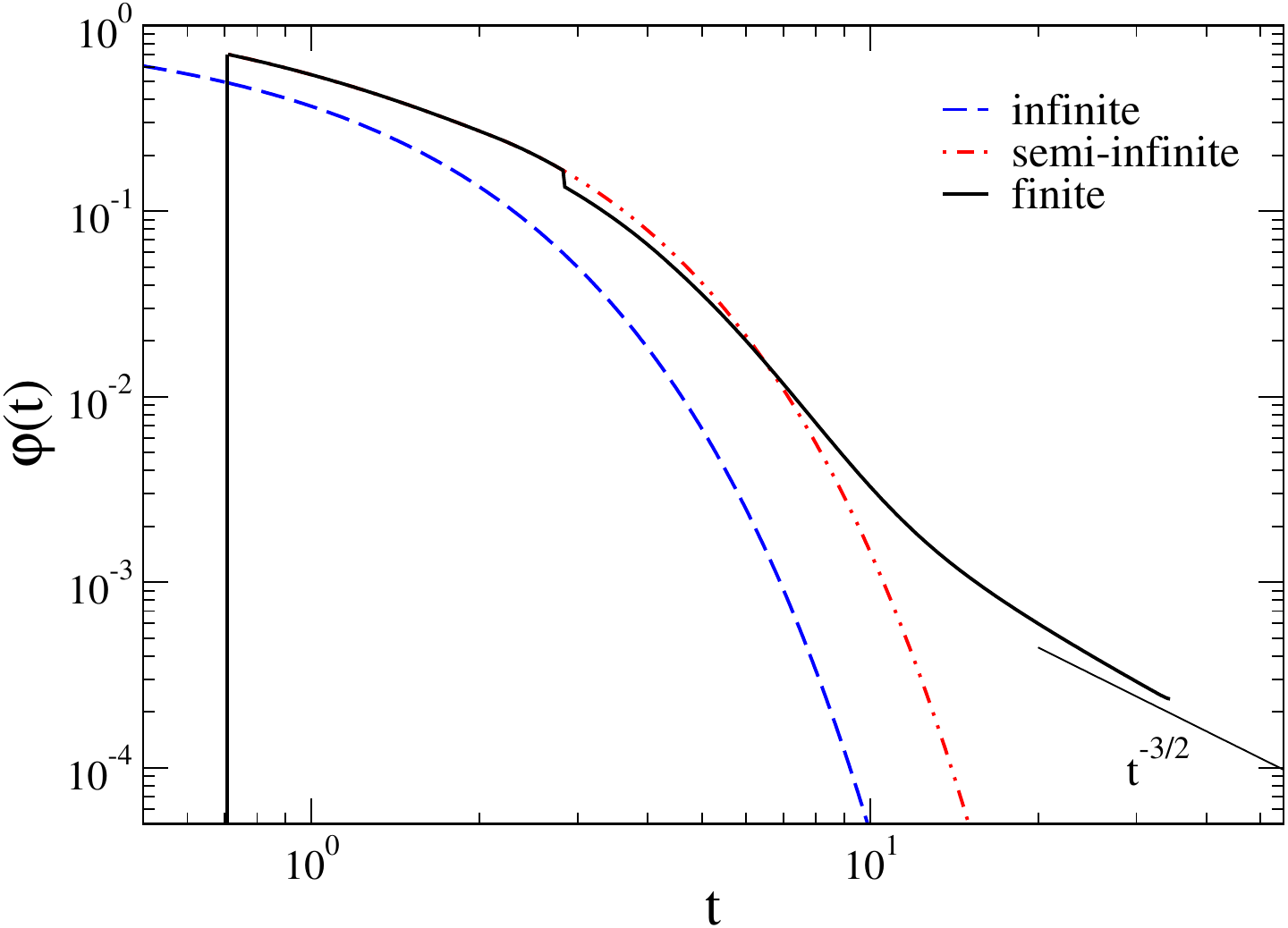}
\caption{\label{fig4}
Trapping time distributions $\varphi(t)$ for the cases of
infinite, semi-infinite and finite extension of trapping regions.
The curve for the infinite case is the exponential (\ref{phi_inf}),
while the curves for the semi-infinite and finite cases are calculated numerically inverting the Laplace transform expressions (\ref{TTDSF}) and (\ref{TTDF}).
The distributions of semi-infinite and finite cases are different from zero only for times
longer that the minimum time $t_a=a/v$
required for the particle to reach the border $x=a$ of the trapping zone, while
the small discontinuous drop present in the finite case corresponds to the first exit 
at  $t_b=b/v$ from 
the outer border $x=b$ of the trapping domain.
We set $\alpha=1$, $v=1$, $\gamma=1$, $a/\lambda=1$ and $b/\lambda=4$ 
($\lambda=1/\sqrt{2}$).
}
\end{figure}
Also in this case we can calculate the mean trapping time.
Let us first study the trapping time distribution.
The survival probability (\ref{SurvP}) is given by 
    \begin{equation}
    \mathbb{P}(t)= 2 
    \left[ 
    \int_0^a dx\ P^{^{({\rm I})}}(x,t) 
    +     \int_a^\infty dx\ P^{^{({\rm II})}}(x,t) \right]
    = 1-2 \int_a^\infty dx\ P_{_B}(x,t) ,
\end{equation}
where we have used normalization condition and the symmetry of the problem.
In the Laplace domain, using the relation
${\tilde {\varphi}}(s) = 1 - s {\tilde {\mathbb{P}}}(s)$ and (\ref{sPB_II}), we obtain the following expression of the trapping time distribution
\begin{equation}
\label{TTDSF}
{\tilde {\varphi}}(s) = 2 \gamma \int_a^\infty 
dx \ {\tilde P}^{^{({\rm II})}}(x,s) =
%\frac{\gamma q c_{_0}^2}{sc} \
%\frac{1}{c \cosh{(c_{_0}a)} + c_{_0} q \sinh{(c_{_0} a)}}.
\frac{\gamma}{s+\gamma} \
\frac{c}{c \cosh{(c_{_0}a)} + c_{_0} q \sinh{(c_{_0} a)}}.
\end{equation}
Some examples of trapping time distributions $\varphi(t)$ are reported in Fig.\ref{fig4}.\\
The mean trapping time is obtained from
$\tau = - \partial_s {\tilde {\varphi}}(s)|_{s=0}$:
\begin{equation}
\label{mttsf}
\tau = \frac{1}{\gamma} + \frac{\alpha a^2}{2v^2} +
\frac{a}{v} \sqrt{\frac{\alpha+\gamma}{\gamma}}.
\end{equation}
This expression is valid for generic particle's properties ($\alpha$, $v$) and environmental
parameters ($\gamma$, $a$). We now discuss some interesting limits.\\
First of all, we note that, for $a=0$, we recover the previous result of infinite trapping regions, $\tau = 1/\gamma$. Instead, for $a\to \infty$ or $\gamma \to 0$, the problem reduces to that of a free
particles in an unbounded domain without trapping, resulting, trivially, in an infinite trapping time.\\
We now analyze the two interesting limiting cases of non-tumbling particles and diffusive particles.
The former is obtained in the limit $\alpha \to 0$ (wave limit) giving rise to
\begin{equation}
\tau_{_{W}} = \frac{1}{\gamma} + \frac{a}{v} ,
\end{equation}
which is, precisely, the sum of the time it takes for the non-tumbling particle to arrive at the 
$a$ boundary of the trapping zone and the average trapping time $1/\gamma$ inside it.\\
The diffusive limit is obtained for 
$\alpha,v \to \infty$ with finite diffusion constant $D=v^2/\alpha$.
In such a case the (Laplace transformed) trapping time distribution reads
\begin{equation}
\label{TTDIFF}
{\tilde {\varphi}}_{_{Diff.}}(s) =
\frac{\gamma}{\sqrt{s+\gamma}} \
\frac{1}{\sqrt{s+\gamma} \cosh{\big(a\sqrt{s/D}\big)} + \sqrt{s} \sinh{\big(a \sqrt{s/D}\big)}} ,
\end{equation}
and the mean trapping time is
\begin{equation}
\tau_{_{Diff.}} = \frac{1}{\gamma} + \frac{a^2}{2D} +
\frac{a}{\sqrt{\gamma D}} .
\end{equation}

\subsection{First-passage problem as a limiting case $\gamma \to \infty$}
Here we show how it is possible to obtain the solution of the free run-and-tumble motion 
in a finite domain $[-a,a]$ with perfectly absorbing boundaries 
taking the limit $\gamma \to \infty$ of the previous results. 
Indeed, in this limit, the particle is instantaneously
absorbed when arriving at the edge $x=\pm a$ of the trapping zone 
and we then get a first-passage problem.
The firs-passage time distribution is then obtained from (\ref{TTDSF}) 
\begin{equation}
\label{FPTD}
{\tilde {\varphi}}(s) =
\frac{1}{\cosh{(c_{_0}a)} + \sqrt{\frac{s}{s+\alpha}} \sinh{(c_{_0} a)}} ,
\hspace{1.3cm} \gamma \to \infty ,
\end{equation}
retrieving previous results in the literature (see Eq.(64) of Ref.\cite{JPA2015}
with $\epsilon=1$). 
The mean first-passage time is obtained from (\ref{mttsf}) and we have \cite{JPA2015,FPT_EPJE} 
\begin{equation}
\tau = \frac{\alpha a^2}{2v^2} +
\frac{a}{v} ,
\hspace{1.3cm} \gamma \to \infty .
\end{equation}
Similarly, we obtain the first-passage time distribution  of a diffusive particle in a finite domain $[-a,a]$ with perfectly absorbing boundaries by taking the limit 
of (\ref{TTDIFF})
\begin{equation}
\label{TTDIFF2}
{\tilde {\varphi}}_{_{Diff.}}(s) =
\frac{1}{\cosh{\big(a\sqrt{s/D}\big)}}  ,
\hspace{1.3cm} \gamma \to \infty ,
\end{equation}
and the mean first-passage time now reads
\begin{equation}
\tau_{_{Diff.}} =  \frac{a^2}{2D} ,
\hspace{1.3cm} \gamma \to \infty .
\end{equation}

%%%%%%%%%%%%%%%%%%%%%%%%%%%%%%%%%%%%%%%%%%%%%%%%%%%

\section{Finite trapping region}
The last case we analyze is that of a finite trapping interval $a<|x|<b$
(see Fig.\ref{fig1}).
We have now to solve three sets of equations, two in the free regions (I) $|x|<a$ and (III) $|x|>b$ with $\gamma=0$, like Eq.s (\ref{PR_I}-\ref{PL_I}), and one in the trapping region (II) $a<|x|<b$
with $\gamma>0$, like Eq.s (\ref{PR_II}-\ref{PB_II}). 
Following similar arguments as in the previous sections, we can write the solutions, in the Laplace domain, as
\begin{eqnarray}
\label{fPA_I}
{\tilde P}^{^{({\rm I})}}(x,s) &=& A^{^{({\rm I})}}_+\ \exp{\left(c_{_0}|x|\right)}
+ A^{^{({\rm I})}}_-\ \exp{\left(-c_{_0}|x|\right)},
\quad \mbox{for $|x|<a$},\\
\label{fPA_II}
{\tilde P}^{^{({\rm II})}}(x,s) &=& A^{^{({\rm II})}}_+\ \exp{\left(c|x|\right)}
+ A^{^{({\rm II})}}_-\ \exp{\left(-c|x|\right)},
\quad \mbox{for $a<|x|<b$},\\
\label{fPA_III}
{\tilde P}^{^{({\rm III})}}(x,s) &=& 
A^{^{({\rm III})}}_-\ \exp{\left(-c_{_0}|x|\right)} ,
\quad \mbox{for $|x|>b$},\\
{\tilde P}_{_B} (x,s) &=&  \frac{\gamma}{s} 
{\tilde P}^{^{({\rm II})}}(x,s) ,
\quad \mbox{for $a<|x|<b$},
\label{fPB_II}
\end{eqnarray}
where $c(s)$ and $c_{_0}(s)$ are given by (\ref{cs}-\ref{c0s}), and 
\begin{eqnarray}
A^{^{({\rm I})}}_{\pm} &=& \mp  \frac{c_{_0}}{4s} \ 
\frac{c k_- \pm c_{_0}qk_+}{c k_- \cosh{(c_{_0}a)} - c_{_0} q k_+ \sinh{(c_{_0} a)}} \ \exp{(\mp c_{_0} a)} , \\
A^{^{({\rm II})}}_{+} &=& -\frac{c_{_0}^2 q}{2s} \ 
\frac{1}{ck_- \cosh{(c_{_0}a)} - c_{_0} q k_+ \sinh{(c_{_0} a)}}
\ \exp{(-c a)} ,
\\
A^{^{({\rm II})}}_{-} &=& \frac{c_{_0}^2 q}{2s} \ 
\frac{1-k_+}{ck_- \cosh{(c_{_0}a)} - c_{_0} q k_+ \sinh{(c_{_0} a)}}
\ \exp{(c a)} ,
\\
A^{^{({\rm III})}}_{-} &=& - \frac{c_{_0}^2 c q}{s(c-c_{_0}q)} \ 
\frac{1}{ck_- \cosh{(c_{_0}a)} - c_{_0} q k_+ \sinh{(c_{_0} a)}}
\ \exp{[c_{_0} b + c(b-a)]} ,
\end{eqnarray}
with $q(s)$ given by (\ref{qs}) and
\begin{equation}
    k_{\pm} = 1 \pm \frac{c+c_{_0}q}{c-c_{_0}q} .
    \ \exp{[2c(b-a)]}
\end{equation}
As before, we can obtain an exact expression for the stationary distribution of blocked particles in the trapping region $a<|x|<b$, obtaining (see Fig.\ref{fig2}) 
\begin{equation}
\label{PBst3}
P_{_B}^{(st.)} (x) = \frac{1}{2\lambda} \ \frac{\cosh{[(b-|x|)/{\lambda}]}}{\sinh{[(b-a)/{\lambda}]}},
\quad \mbox{for $a<|x|<b$},
\end{equation}
where $\lambda$ is the characteristic length (\ref{lamb}).
We note that for $b\to\infty$ we recover the previous semi-infinite case (\ref{PBst2}).\\
We now study the distribution of trapping time.
The survival probability is given by 
    \begin{equation}
    \mathbb{P}(t) = 1-2 \int_a^b dx\ P_{_B}(x,t) .
\label{phit3}
\end{equation}
By using the expressions (\ref{fPB_II}) and (\ref{fPA_II}) and the relation
${\tilde {\varphi}}(s) = 1 - s {\tilde {\mathbb{P}}}(s)$,
we have that the trapping time distribution in the Laplace domain reads
\begin{equation}
\label{TTDF}
{\tilde {\varphi}}(s) = 
%2 \gamma \int_a^\infty dx \ {\tilde P}^{^{({\rm II})}}(x,s) =
%\frac{\gamma q c_{_0}^2}{sc} \
%\frac{(1-e^{-c(b-a)})\ (1-k_+-e^{c(b-a)}) }
\frac{\gamma}{s+\gamma} \
\frac{c\ (1-e^{-c(b-a)})\ (1-k_+-e^{c(b-a)}) }
{ck_- \cosh{(c_{_0}a)} - c_{_0} q k_+ \sinh{(c_{_0} a)}} ,
\end{equation}
where $L=b-a$.
For small $s$ we have that 
\begin{equation}
{\tilde {\varphi}}(s) \sim 1 - A \sqrt{s} -B s ,
\label{phif}
\end{equation}
with prefactors $A$ and $B$ depending on the system and geometrical parameters. We have:
\begin{equation}
A = \sqrt{\frac{\alpha+\gamma}{\alpha \gamma}} \ 
\frac{1}{\sinh{(L/\lambda)}} ,
\label{termA}
\end{equation}
and
\begin{equation}
    B = \frac{1}{\gamma} + \frac{\alpha a^2}{2v^2} +
    \frac{\cosh{(L/\lambda)}}{\sinh{(L/\lambda)}}
    \Bigg( \frac{a}{v} \sqrt{\frac{\alpha + \gamma}{\gamma}} - 
    \frac{\alpha+\gamma}{\alpha \gamma}\ 
    \frac{1}{\sinh{(L/\lambda)}}   \Bigg).
\label{termb}
\end{equation}
At small $s$ the survival probability 
diverges as
${\tilde{\mathbb{P}}}(s) \sim s^{-1/2}$.
From the Tauberian theoremes we have that, in time domain, ${\mathbb{P}}(t) \sim t^{-1/2}$ for large $t$
\cite{ksbook}.
This implies that the trapping time distribution,
given by $\varphi(t) = -\partial_t {\mathbb{P}}(t)$,
behaves asymptotically  as
\begin{equation}
\varphi(t) \sim  \frac{A}{2\Gamma(1/2)}\  t^{-3/2},
\hspace{1.5cm} t \to \infty,
\end{equation}
where $\Gamma(x)$ is the Gamma function (see Fig.\ref{fig4}).
%(see Eq. (3.19) in \cite{ksbook}). 
We then conclude that the mean trapping time (\ref{MTT})  diverges, as a consequence of the infinite extension of the free zone, resulting in a slower trapping of particles.
It is worth noting that, in the limit of semi-infinite trapping region ($b \to \infty$) the prefactor $A$ in (\ref{phif}) vanishes
(see (\ref{termA}) for $L \to \infty$)
and Eq.(\ref{TTDF}) reduces to (\ref{TTDSF}).
The mean trapping time is then finite and it is given by the term $B$ in (\ref{termb}) for $L \to \infty$, that coincides with the expression (\ref{mttsf}). 
Finally, we note that the case of a particle starting its motion at the center of a finite trapping zone $[-b,b]$ is simply obtained 
by taking the limit $a \to 0$ of the previous results.

%%%%%%%%%%%%%%%%%%%%%%%%%%%%%%%%%%%
\section{Modeling anomalous diffusion in the trapping region}

\begin{figure}[t!]
\includegraphics[width=.7\linewidth] {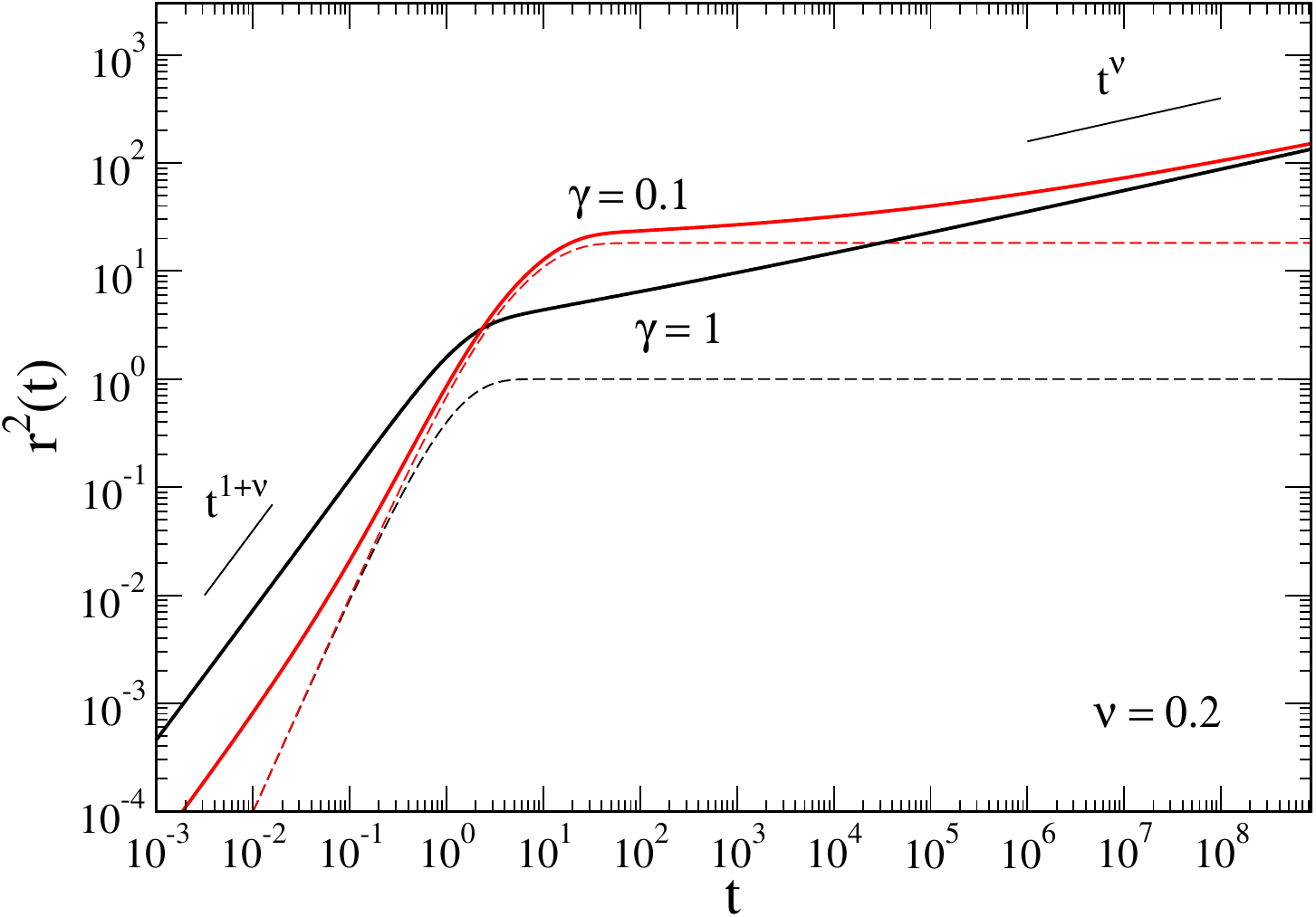}
\caption{\label{fig5}
Mean-square displacement $r^2(t)$ for the fractional diffusion model in the case of infinite extension of the
trapping region. We consider fractional derivative exponent $\nu=0.2$. 
Black and red continuous lines correspond, respectively, to trapping rates $\gamma=1$ and $\gamma=0.1$.
Dashed lines are the corresponding curves in the case of complete blocking of particles, i.e., $D=0$ 
(see section \ref{sec3}).
We set $\alpha=1$, $v=1$, $D=1$.
}
\end{figure}

Bacterial adhesion to surfaces often occurs through complex and nontrivial mechanisms.
For example in \cite{Sjo} it was shown that the cell adhesion to glass surfaces
involves  multiple reversibly-binding tethers that detach and successively re-attach,
resulting in a slowing down of the dynamics of attached bacteria. 
The mean-square displacement of 
several bacterial strains was found to have a subdiffusive trend at long times,
$r^2 \sim t^\nu$ with $\nu<1$ \cite{Sjo}.
In this last section we extend our model to take into account in an effective way such a subdiffusive
character of the bacterial dynamics in the trapping regions.
To this end, we make use of fractional diffusion models, 
which are known to generate subdiffusive dynamics at long times 
\cite{ksbook,comp,MK2000,Ran2000,Maso,Anga2020}.
For the sake of simplicity, here we analyze only the case of infinitely extended 
trapping zone (see section \ref{sec3}).
The model describes a run-and-tumble particle that, at fixed rate, irreversibly switches to 
a phase characterized by anomalous diffusion. 
By introducing the time-fractional derivative of order $\nu \in (0,1)$, 
\begin{equation}
    \frac{\partial^\nu f}{\partial t^\nu} (x,t) = I^{1-\nu} \frac{\partial f}{\partial t} (x,t) , 
\end{equation}
where $I^\mu$ is the Riemann-Liouville fractional intergral
\begin{equation}
%    I^{1-\nu} f(x,t) = \frac{1}{\Gamma(1-\nu)} \int_0^t d\tau (t-\tau)^{-\nu} f(x,\tau) ,
   I^{\mu} f(x,t) = \frac{1}{\Gamma(\mu)} \int_0^t d\tau (t-\tau)^{\mu-1} f(x,\tau) ,
       \hspace{1.3cm} \mu>0 ,
\end{equation}
we can generalize the equation (\ref{eq_PB}) for the PDF of particles in the trapping regions 
including fractional diffusion in the following manner
\begin{equation}
\label{PDF_fr}
    I^{1-\nu} \left( \frac{\partial P_B}{\partial t} - \gamma P \right)
    = D \frac{\partial^2 P_B}{\partial x^2} .
\end{equation}
The equations for $P$ and $J$ are the same as (\ref{eq_PA}) and (\ref{eq_J}).
We note that in the case of null diffusion, $D=0$, we recover the original case
of immobilized particles, eq. (\ref{eq_PB}), as $I^{1-\nu} f = 0$ implies $f=0$.
Instead, in the limit $\nu \to 1$, we have that $I^{1-\nu} f \to f$, and we obtain
the case of normal diffusion
\begin{equation}
    \frac{\partial P_B}{\partial t} =
    D \frac{\partial^2 P_B}{\partial x^2} + \gamma P ,
   \hspace{1.3cm} \nu \to 1 .
\end{equation}
Proceeding as in section \ref{sec3}, we can write the eq. (\ref{PDF_fr}) in the Laplace domain
as
\begin{equation}
\label{Pfr_L}
    s^\nu {\tilde P}_{_B} = D \frac{\partial^2 {\tilde P}_{_B}}{\partial x^2}
   + \gamma s^{\nu-1} {\tilde P} ,
\end{equation}
having used the fact that 
\begin{equation}
    \mathcal{L} [I^{1-\nu} f(t) ] (s) = \frac{{\tilde f}(s)}{s^{1-\nu}} .
\end{equation}
Performing now the Fourier transform 
\begin{equation}
    \hat{f}(k) \equiv \mathcal{F} [f(x)](k) = \int_{-\infty}^{+\infty} dx \ e^{ikx} f(x) , 
\end{equation}
the equation (\ref{Pfr_L}) becomes
\begin{equation}
\label{Pfr_LF}
    (s^\nu + D k^2) {\hat {\tilde{P}}}_{_B} = \gamma s^{\nu-1} {\hat {\tilde P}}  .
\end{equation}
The PDF in the RHS can be obtained from eq. (\ref{teleg})
(we remind that the equation for $P$ is the same as in section \ref{sec3})
leading to
\begin{equation}
\label{Pks}
    {\hat {\tilde P}} = \frac{s + \gamma + \alpha}
    {(s+\gamma)(s+\gamma+\alpha) + v^2 k^2} .
\end{equation}
We have then obtained the exact expressions of $P$, eq. (\ref{Pks}) and $P_B$, from eq. (\ref{Pfr_LF}),
in the Laplace-Fourier domain,
thus allowing us to compute the mean-square displacement thorough
\begin{equation}
    {\tilde {r^2}} (s) = - \left. \frac{\partial^2}{\partial k^2} 
    (  {\hat {\tilde P}} + {\hat {\tilde{P}}}_{_B}) \right|_{k=0} .
\end{equation}
After some algebra we finally obtain
\begin{equation}
\label{r2fLap}
    {\tilde{r^2}}(s) = 
    \frac{2 v^2}{s (s+\gamma)(s+\gamma+\alpha)} +
    \frac{2\gamma D}{s^{\nu+1}(s+\gamma)} \equiv {\tilde{r_{_A}^2}}(s) + {\tilde{r_{_B}^2}}(s)  ,
\end{equation}
which generalizes eq. (\ref{r2Lap}) to the present case of fractional diffusion.
With respect to the original expression, we note here the presence of a second term, 
which takes into account the anomalous diffusion of trapped particles.
The mean-square displacement in the time domain is obtained 
by performing the inverse-Laplace transform of the previous expression
\begin{equation}
\label{rAB2}
    r^2(t) = r_{_A}^2(t) + r_{_B}^2(t) ,
\end{equation}
with the first term that is the same 
obtained in section \ref{sec3}, eq. (\ref{r2}),
\begin{equation}
r_{_A}^2(t) = \frac{2v^2}{\alpha \gamma (\alpha+\gamma)} 
\left[ 
\alpha (1-e^{-\gamma t}) - 
\gamma e^{-\gamma t } (1-e^{-\alpha t})
\right] ,
\label{rA2}
\end{equation}
and the second term that can be expressed as 
\begin{eqnarray}
\label{rB2}
 r_{_B}^2(t) &=&   \frac{2 \gamma D}{\Gamma(2+\nu)}\  t^{1+\nu} \ e^{-\gamma t} \ 
 \Phi(\nu+1,\nu+2;\gamma t) \nonumber \\
 &=& \frac{2 \gamma D}{\Gamma(2+\nu)}\  t^{1+\nu}\   \Phi(1,\nu+2;- \gamma t) ,
\end{eqnarray}
where we have introduced the degenerate (confluent) hypergeometric function \cite{TISP_Book}
\begin{equation}
    \Phi(\beta,\mu;z) = \frac{\Gamma(\mu)}{\Gamma(\beta)\Gamma(\mu-\beta)}\ z^{1-\mu} \ 
    \int_0^z dt\  e^t\  t^{\beta-1} \ (z-t)^{\mu-\beta-1} \equiv {_1F_1}(\beta;\mu;z) ,
\end{equation}
with $\Gamma(z)= \int_0^\infty t^{z-1} \ e^{-t}\ dt$ the Euler Gamma function,
and we have used the property
\begin{equation}
    \Phi(\beta,\mu;z) = e^z\ \Phi(\mu-\beta,\mu;-z) .
\end{equation}
Let us analyze the asymptotic behaviors.
In the long time regime the dominant terms in eq. (\ref{r2fLap}) are 
obtained for small  $s$, 
\begin{equation}
{\tilde{r^2}}(s) \sim \frac{2D}{s^{\nu+1}} + \frac{2v^2}{\gamma(\gamma+s)} \ \frac1s ,
       \hspace{1.3cm} s \to 0 ,
\end{equation}
corresponding in the time domain to
%an anomalous diffusion behavior
\begin{equation}
    r^2(t) \sim  
    \frac{2D}{\Gamma(\nu +1)} \ t^\nu +
    \frac{2v^2}{\gamma(\gamma+\alpha)} 
    \sim \frac{2D}{\Gamma(\nu +1)} \ t^\nu ,
% \hspace{1.3cm} \textrm{for large t}   
       \hspace{1.3cm} t \to \infty .
\end{equation}
We then obtain, asymptotically, anomalous diffusion with exponent $\nu$.
We note that, for immobilized particles, $D=0$, the dominant term is the constant one, 
and the MSD develops a plateau (\ref{r2plat}).
The same asymptotic trend can be inferred directly from the expression (\ref{rB2}).
Indeed, by manipulating the integral in the hypergeometric function, we can 
write  (\ref{rB2}) in the form
\begin{equation}
    r_B^2(t) = \frac{2\gamma D}{\Gamma(\nu+1)} \ t^\nu \ 
    \int_0^t d\tau \ e^{-\gamma \tau} (1-\tau/t)^\nu 
    \sim \frac{2D}{\Gamma(\nu +1)} \ t^\nu ,
       \hspace{1.3cm} t \to \infty ,
\end{equation}
as the integral converges to $1/\gamma$ in the asymptotic limit.\\
At short times the dominant terms are obtained for large $s$,
\begin{equation}
{\tilde{r^2}}(s) \sim \frac{2\gamma D}{s^{\nu+2}} + \frac{2v^2}{s^3} ,
       \hspace{1.3cm} s \to \infty ,
\end{equation}
corresponding to 
\begin{equation}
    r^2(t) \sim 
    \frac{2\gamma D}{\Gamma(\nu +2)} \ t^{\nu+1} +
    v^2 \ t^2 
    \sim \frac{2\gamma D}{\Gamma(\nu +2)} \ t^{\nu+1}  ,
       \hspace{1.3cm} t \to 0 .
\end{equation}
In figure \ref{fig5} we report the mean-square displacement 
(\ref{rAB2}) for the case of exponent $\nu=0.2$ (close to  typical values
obtained in experiments \cite{Sjo}).
The curves (full lines) correspond to two different values of the trapping rate,
$\gamma=1$ and $\gamma=0.1$. For comparison we report also the corresponding cases
of immobilized particles in the trapping zone, $D=0$ (dashed lines), studied in 
section \ref{sec3}.
It is evident an anomalous subdiffusive behavior $t^\nu$ at long time, a  superdiffusive behavior 
$t^{\nu+1}$ at short time and 
 possible   intermediate regimes (ballistic, diffusive or plateau-like) depending
on the parameters values.

%%%  Figure  %%%%%%%%%%%%%

%%%%%%%%%%%%%%%%%%%%%%%%%%%%%%%%%%%
\section{Conclusions}
In this work we studied the problem of active particles in trapping environments,
describing the irreversible adhesion processes that take place, for example, in the early stage of biofilms formation.
In particular, we considered 1D run-and-tumble particles in the presence of trapping regions where particles absorption takes place at rate $\gamma$. Different extension of the trapping regions are investigated: infinite, semi-infinite and finite.
By solving the kinetic equations for the probability density functions we are able to provide exact expressions of several interesting quantities.
The case of infinite trapping interval is fully solvable in the time domain. We report expressions of PDFs of moving and blocked particles, mean-square displacement,
trapping time distribution and mean trapping time, which turns out to be simply the inverse of
the trapping rate $1/\gamma$.
In the case of semi-infinite trapping region we are able to solve the problem in the Laplace domain, allowing us to give exact expressions of stationary distribution of blocked particles (\ref{PBst2}) 
and mean trapping time (\ref{mttsf}). Several limiting cases are also analyzed, such as 
diffusive motions and first-passage problems in a finite domain.
Finally, we analyze the case of a finite trapping region, reporting again the spatial distribution of blocked particles in the stationary regime (\ref{PBst3}) and discussing the behavior of the trapping time distribution, whose long tail at large $t$ leads to divergent mean trapping time.
A last section is devoted to extend the model to the case of anomalous diffusion of trapped particles,
in  accordance with some experimental observations on bacterial adhesion to glass surfaces \cite{Sjo}.
By resorting to fractional diffusion models we are able to derive exact expressions of the MSD,
resulting in subdiffusive behaviors $r^2 \sim t^\nu$ with $\nu<1$ in the long time regime 
and non-trivial trends in the intermediate regimes.
\\
It would be of interest to extend the present analysis 
in different directions.
A first implementation might be to consider
reversible trapping, that is, the possibility for the particle to reactivate itself after 
%being immobilized 
trapping \cite{Per2023}.
Other possible extensions could be the analysis of planar motions
\cite{mart,Ang2022,basu,smith,Ors}
or considering more complex environments,
%spatial extension of trapping zones
such as those described by a continuously variable trapping rate, 
by a periodic sequence of trapping intervals \cite{Poz2021,Poz2022}
or by the presence of generic boundaries \cite{JPA_2023}.
A final possible direction of investigation might be to consider different combinations
of particle motion in the two phases, before and after trapping.
In the present study we investigated the case of active motion before
trapping and arrested phase or anomalous diffusive phase after it.
It would be interesting to consider also, for the initial active phase,
fractional processes \cite{Maso,Anga2020,Dean2021} or generalized $g$-fractional motions
 \cite{Anga2023,AG2023},
extending, for example, the recent investigation on subdiffusive processes with 
particles immobilization \cite{Kos2023}. 
A final remark concerns the modeling of the entire biofilm formation process.
It could be of great importance to implement the described run-and-tumble models in more realistic contexts (also using numerical simulations), considering both irreversible adhesion and 
cellular replication \cite{PAM}.
%}

%%%%%%%%%%%%%%%%%%%%%%%%%%%%%%%%%%%
\acknowledgments
%\paragraph*{Acknowledgments.}
I thank Roberto Garra for useful discussions.
I acknowledge financial support from the Italian Ministry of University and Research (MUR) under 
PRIN2020 Grant No. 2020PFCXPE.

%%%%%%%%%%%%%%%%%%%%%%%%%%%%%%%%%%%
%%%%%%%%%%%%%%%%%%%%%%%%%%%%%

\end{document}